# Application of Mythen Detector: In-situ XRD Study on The Thermal Expansion Behavior of Metal Indium*

DU Rong (杜蓉), CHEN Zhongjun (陈中军), CAI Quan (蔡泉),

FU Jianlong (傅建龙), GONG Yu (宫宇), WU Zhonghua (吴忠华) [1)]

Institute of High Energy Physics and University of the Chinese Academy of Sciences,
Chinese Academy of Sciences, Beijing 100049, China

**Abstract:**    A Mythen detector has been equipped at the beamline 4B9A of Beijing Synchrotron Radiation Facility, which can be used for in-situ real-time measurement of X-ray diffraction (XRD) full profiles. In this paper, the thermal expansion behavior of metal indium has been studied by using the in-situ XRD technique with the Mythen detector. The indium film was heated from 30 to 160 ℃ with a heating rate of 2 ℃/min. The in-situ XRD full-profiles were collected with a rate of one profile per 10 seconds. Rietveld refinement was used to extract the structural parameters. The results demonstrate that the thermal expansion of metal indium is nonlinear especially when the sample temperature was close to its melting point (156.5 ℃). The expansion of $a$-axis and the contraction of $c$-axis of the tetragonal unit cell of metallic indium can be well described by biquadratic and cubic polynomials, respectively. The tetragonal unit cell presents a tendency to become cubic one with the increase of temperature but without detectable phase change. This study is not only beneficial to the application of metal indium, but also exhibits the capacity of in-situ time-resolved XRD experiments at the X-ray diffraction station of BSRF.

## 1   Introduction

Thermal expansion coefficient of metal materials is an important parameter for metal application and their thermal stability. Therefore, the study on thermal expansion behavior of metal materials has not only great academic interests, but also application value. In-situ temperature-dependent X-ray diffraction (XRD) is the main experimental method to get the precise thermal expansion coefficient. In the conventional measurements, the metal samples [1-2] were usually heated to the wanted target temperatures and kept at these temperatures for enough long time to reach the thermal equilibrium, then the corresponding XRD patterns were collected by a step-scanning mode. Afterwards, the thermal expansion coefficient can be extracted by analyzing the change of lattice parameters from the XRD patterns. Generally, such a conventional XRD mode is quite time-consuming. Each collection of XRD patterns could span several hours or more. However, the crystalline structure or intermediate products could be fast changed for a dynamic process or chemical reaction, for example, the appearance of new species or the disappearance of old species in chemical reaction, the fast evolution of crystallographic structure around phase transformation point, the drastic increase of structural disorder near the melting point. In such cases, time-resolved XRD technique is necessary for tracking the structural changes with the heating temperature or the reaction time. Partial time-resolved XRD measurements [3-5] have been reported by collecting the XRD patterns with two-dimension (2D) detector like CCD or image plate. The time resolution of those measurements depends on the response speed of the 2D detector and the incident X-ray intensity. It is well known that the response speed of 2D detectors is about 2.5 s for a CCD detector or more for an image plate detector, thus a best time resolution could be achieved in second level. However, these 2D detectors also have obvious disadvantage as used in a time-resolved technique. Due to the limited active area of the 2D detectors, the range of diffraction angles collected by a CDD or an image plate detector is quite restricted. Therefore, it is difficult to collect a full pattern of the sample so that the recorded XRD data cannot be used for structural refinement.

Recently, a microstrip detector system [6] named Mythen detector has been equipped at the XRD station of beamline 4B9A of Beijing Synchrotron Radiation Facility (BSRF). This detector consists of 24 detection modules and can be used for time-resolved XRD measurements in millisecond level. The 24 modules are assembled on an install circle with radius of about 760 mm, forming a curved detector. The covering angle (2θ) of this detector is 120° with total 30720 pixels. Physically, each module contains 1280 pixels, each pixel corresponds to 0.00377°, and a small blind-gap is left between two adjacent modules. It is important that the readout time of single module is about 0.3 ms. The total readout time of the whole detector is also in millisecond level, depending on the data bit depth. Therefore, this detector is promised to perform in-situ, real-time, and full-profile XRD measurement with time resolution in sub-second level.

Metal indium or indium-gallium alloys are often used as heat contact medium between optical element and cooling body due to their flexibility. The anisotropic structure of metal indium with tetragonal symmetry was reported [7-9] to have different thermal expansion coefficients in the crystallographic orientations of *a*- and *c*-axes. For a better application of metal indium as a heat contact medium, the knowledge about its thermal expansion and heat stability is desired. In this paper, the thermal expansion behavior of metal indium is studied by using the Mythen detector.

## 2  Experimental Details

Indium film with purity of 99.99% was used as the sample. In-situ XRD measurements changed with heating temperature were performed at the XRD station of beamline 4B9A of BSRF. The incident X-ray wavelength was chosen to be 1.54 Å. The XRD patterns at different heating temperatures were collected with the Mythen detector. In these measurements, the indium film was sealed by Kapton film and put onto the copper basement as shown in Fig. 1. The copper basement was set to have a tilt angle of 10° related to the incident X-ray beam, and was used to heat the indium film from the bottom. The sample was heated to temperature ranging from 30 to 160 ℃ with a heating rate of 2 ℃/min. The in-situ XRD patterns were continuously recorded by the Mythen detector with an exposure time of 10 seconds for each pattern. Thus, each XRD pattern is corresponding to a temperature span of 1/3 ℃. The collected diffraction intensity (*I*) curves versus channel (or pixel) number (*CH*) of the Mythen detector were first converted into the diffraction-intensity (*I*) curves versus diffraction angle (2θ). Simultaneously the angular calibration and flat-filed correction of the detector were also performed with a data pretreatment program. A detailed description about the data pretreatment can be found elsewhere [10].

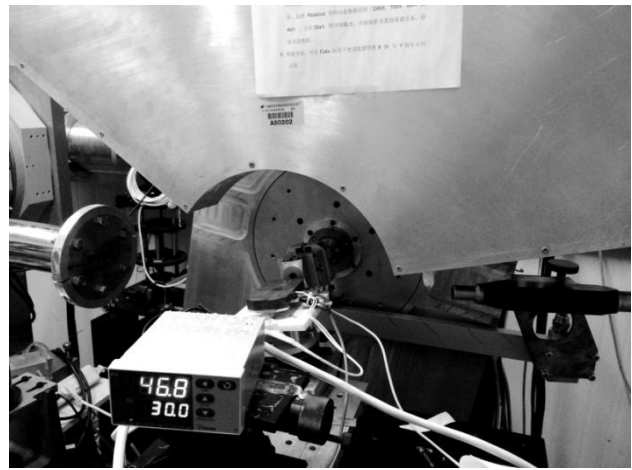

**Figure 1**. Photo of indium film, copper basement, and the Mythen detector. The indium film is sealed with Kapton film and put onto the copper basement which can be used to heat the sample from room temperature to 160 ℃.

## 3   Result and Discussion

After the data pretreatment, the final XRD patterns were obtained. Partial representative XRD patterns of the indium film changed with heating temperature are shown in Fig. 2(a). It can be seen that only few diffraction peaks appear in the angular range from 25° to 115°, but there is an intense diffraction peak appearing around 56°. To further determine the structural change of the indium film with temperature, Jade 6.5 program [??] was firstly used to index the diffraction peaks. The result demonstrates that (200) reflection has the strongest intensity at lower temperatures, which implies that the initial indium film has a preferred orientation along [200] direction. It is well known that metal is easily oxidized in atmosphere. We believe that there must be an oxidation layer on the surface of the indium film because the sample was not surface-treated before use. However, no peaks from indium oxide can be found from the XRD patterns. On the one hand, this result implies that the oxidation layer on the surface of the initial indium film is so thin that it can be ignored. On the other hand, it also illustrates that the indium film has been well protected from oxidation by sealing the indium film with Kapton film. The melting point of metal indium was reported [11] to be 156.5 °C. These diffraction patterns near the melting point of indium are compared in Fig. 2(b). It can be found that all diffraction peaks are completely disappeared as the sample temperature is higher than 156.5 °C, which clearly tells us that the melting point of the indium film is 156.5 °C. This result is excellently agreement with the existing report [11], which also proves indirectly that the surface oxidation of indium film has no detectable effect on the melting point of metal indium. Even the surface oxidation layer is beneficial to the protection of the indium from further oxidation.

It has been known that the crystalline structure [12] of metal indium is a tetragonal structure. The space group of crystalline indium is I4/mmm. There are two equivalent atoms in the unit cell with coordinates of (0, 0, 0) and (1/2, 1/2, 1/2). Based on the known space group and the specific atom coordinates, accurate lattice parameters of metal indium can be extracted by refining the XRD patterns with the Rietveld refinement method. Due to a flat sample with tilt angle of 10 ° was used in the in-situ XRD measurements, an absorption correction [13] of the flat sample has been considered in the diffraction intensity refinement. Besides, the strong (200) preferred orientation of the indium film as shown in Fig. 2 suggests that the texture had to be considered in the structural refinement. The Maud program [14] was used for the structural refinements, in which several selectable models for texture analysis are available. In this study, the March-Dollase model [15] was adopted. As an example, the Rietveld refinement for the diffraction pattern of sample at room temperature is shown in Fig. 3.

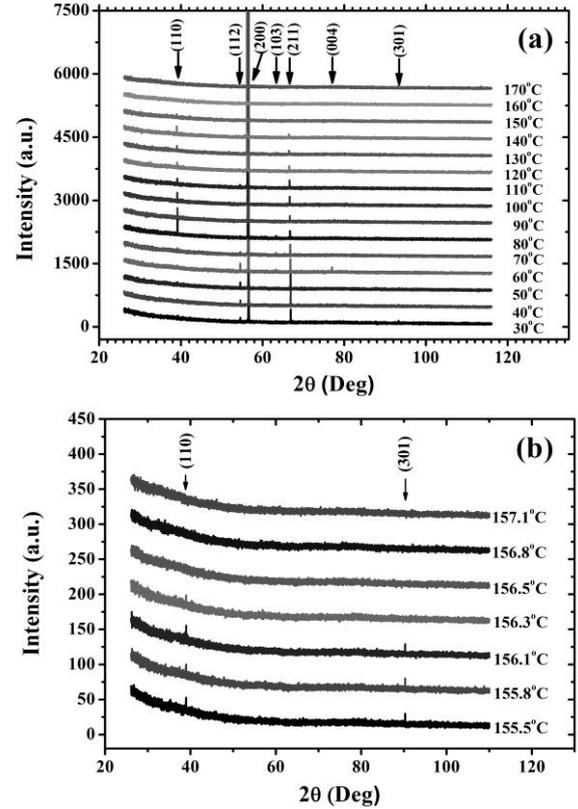

**Figure 2.** (a) Partial representative XRD patterns of the indium film with temperature from 30 to 170 ℃; (b) Several XRD patterns around the melting point of the indium film.

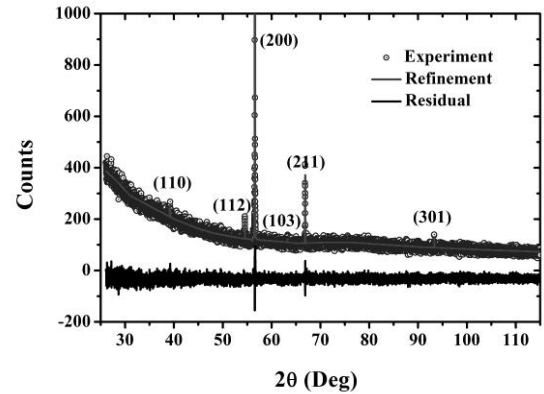

**Figure 3.** Rietveld refinement for the XRD pattern of the indium film at room temperature.

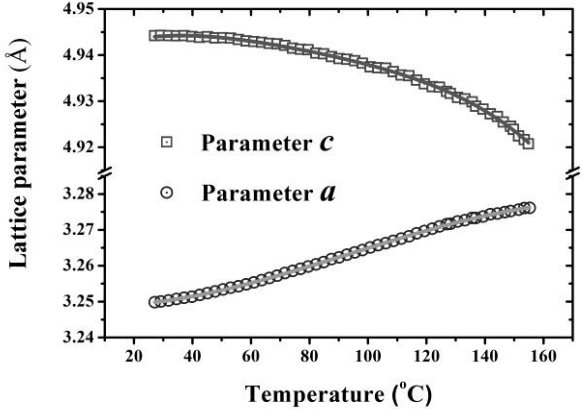

**Figure 4.** Variation of lattice parameters *a* (a) and *c* (b) with temperature.

Based on the above structure refinements, the lattice parameters *a* and *c* are obtained as shown in Fig. 4. It can be found that the lattice parameter *a* increases but the lattice parameter *c* decreases with temperature increasing. Evidently, both changes of parameters *a* and *c* with temperature are all not linear. A biquadratic polynomial as listed in Eq. (1) has to be used to describe accurately the changes of lattice parameters with temperature in degree centigrade.

$$a(T)[\text{or } c(T)] = b_0 + b_1T + b_2T^2 + b_3T^3 + b_4T^4. \tag{1}$$

The fitting curves to Eq. (1) are also shown in Fig. 4 as the solid lines, and the fitting coefficients are listed in Table I.

**Table I.** Fitting coefficients of polynomial

|   | $b_0$ | $b_1(\times10^{-4})$ | $b_2(\times10^{-6})$ | $b_3(\times10^{-8})$ | $b_4(\times10^{-10})$ |
|---|---|---|---|---|---|
| *a* | 3.249 | -0.584 | 3.345 | -1.191 | 0 |
| *c* | 4.939 | 3.546 | -7.495 | 5.622 | -1.770 |

From the temperature-dependent lattice parameters, the coefficient of thermal expansion can be obtained as expressed in Eq. (2).

$$\alpha(T) = \frac{1}{L_0}\frac{dL(T)}{dT}. \tag{2}$$

Where, $L(T)$ is the lattice parameter at temperature T, $L_0$ is the initial lattice parameter at room temperature. The obtained coefficients ($\alpha_a$, $\alpha_c$) of thermal expansion from the measured values are shown in Fig. 5 as the hollow scattered points. It can be seen that the coefficients of thermal expansion are indeed not constants either for parameters *a* or *c*, confirming that the thermal expansion behavior of metal indium is not linear. A previous report [16] about the thermal expansion of metal indium is also compared in Fig. 5 as the solid scattered points, where only three temperature points were measured. Although the previous coefficients of thermal expansion along the *a*- and *c*-orientations were not also constant, their changes with temperature are linear. Combing Eq. (1) and (2), the coefficients of thermal expansion can be calculated as shown in Fig. 5 as the solid lines. The corresponding coefficients of thermal expansion along the *a*-orientation and *c*-orientations of metal indium can be described as Eq. (3) and (4), respectively.

$$10^{10}\alpha_a = -179764+20590.6T-110.010T^2, \tag{3}$$

$$10^{10}\alpha_c = 717953-30350.1T+341.517T^2-1.43320T^3. \tag{4}$$

Unquestionably, these measurement values are consistent with calculated ones. Although this study and the previous report [16] all agree the nonlinear thermal expansion behavior of metal indium, their difference on the thermal expansion is still non-ignorable. The previous report exhibited quadratic temperature dependencies for both parameters *a* and *c*, but this study reveals a biquadratic temperature dependency for the lattice parameter *c* or a cubic one for the lattice parameter *a*. This difference can be attributed to a few of measured points and the absence of higher temperature data near the melting point of metal indium in the previous report. Of course, a linear coefficient of thermal expansion is more convenient to evaluate the lattice parameters at different temperatures. First, the average values of the coefficients of thermal expansion shown in Fig. 4 are obtained. That is: $\widetilde{\alpha}_a = 6.08\times10^{-5}$ and $\widetilde{\alpha}_c = -4.28\times10^{-5}$. It is important to be aware that the lattice parameters $a$ and $c$ evaluated with the above two average coefficients ($\widetilde{\alpha_a}$, $\widetilde{\alpha_c}$) diverge more apparently from the experimental values when the temperature is near to the melting point. The maximum deviation of lattice parameter $c$ occurs at 155 ℃ and is about 0.0180 Å, which corresponds to a relative deviation of 0.366%. The maximum deviation of parameter $a$ also occurs at 155 ℃ and is about -0.0185 Å, which corresponds to a relative deviation of -0.366%. Another liner approximation is closer to the measured values, that is direct fitting to the experimental values of lattice parameters $a$ and $c$ with a linear equation. In this case, the approximative linear coefficients of thermal expansion are obtained to be $2.08\times10^{-4}$ for parameter $a$ and $-1.28\times10^{-4}$ for parameter $c$, respectively. Using the two approximate linear coefficients of thermal expansion, the deviation of the evaluated lattice parameter $a$ from the experimental values locates at a range from minimum -0.0010 Å at 127 ℃ to maximum 0.0015 Å at 61 ℃, and the deviation of the evaluated lattice parameter $c$ from the experimental values locates at a range from minimum -0.0037 Å at 80 ℃ to maximum 0.0071 Å at 155 ℃.

Based on the measured lattice parameters, the volume of unit cell and the axial ratio of $c/a$ were also calculated as shown in Fig. 6. It can be seen that the unit cell volume increases quickly with the temperature increasing from room temperature to about 130 ℃, but it experiences a slower change after 130 ℃. This slower change of unit cell volume implies probably the disintegration of bulk indium and approaching melting. However, the axial ratio $c/a$ is always decreased with the temperature up to the melting point. Generally, the tetragonal unit cell of metal indium has a tendency changed toward cubic one with the increase of temperature. The coefficient of the volume thermal expansion was also obtained as shown in Fig. 7 as the hollow scattered points. A previous report [16] about the coefficient of volume thermal expansion is also shown in Fig. 7 as the solid scattered points. Once again, the coefficient of volume thermal expansion is not a constant and can be described by a quadratic polynomial as formulated as Eq. (5) and shown in Fig. 7 as a solid line.

$$10^5\alpha_v = -3.12008 + 0.40223\times T - 0.00252\times T^2 \qquad (5)$$

Moreover, an empirical formula [17] as formulated as Eq. (6) is also frequently used to evaluate the coefficient of volume thermal expansion.

$$\alpha = \frac{0.020}{T_m} \qquad (6)$$

Where $T_m$ is the melting point of metals in absolute temperature. The empirical coefficient of volume thermal expansion is also shown in Fig. 7 as the dash line. At lower temperatures, these coefficients of volume thermal expansion are roughly consistent, but this study also gives the high-temperature values.

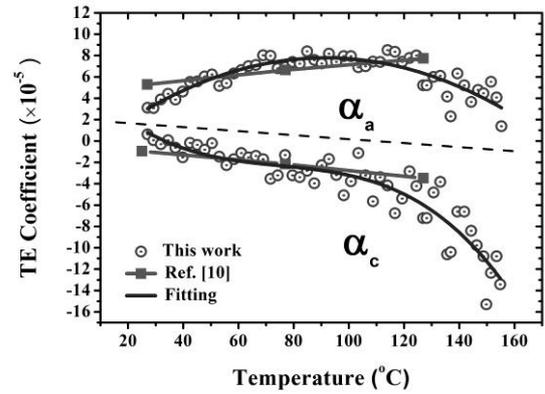

**Figure 5.** Coefficients of thermal expansion of metal indium in $a$-orientation and $c$-orientation. The hollow scattered points are the experimental values, solid scattered points are from Ref. [16], and the solid lines are the fitting polynomials.

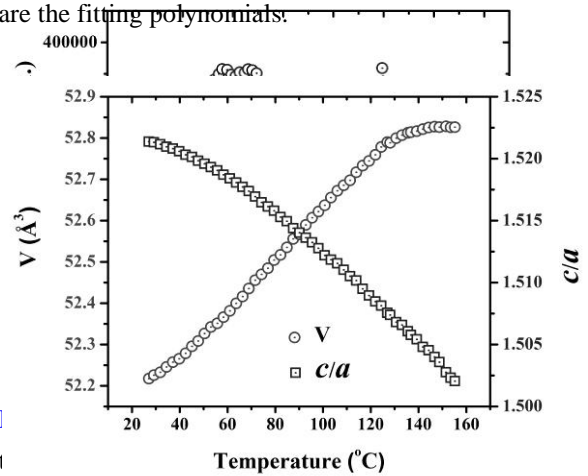

**Figure 6.** Variations of unit cell volume and axial ratio $c/a$ with temperature.

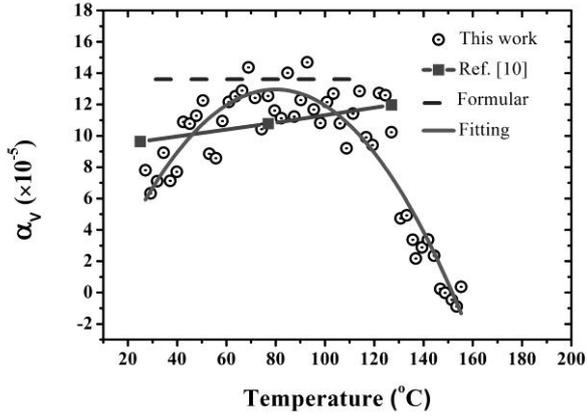

**Figure 7.** Coefficient of volume thermal expansion. Hollow scattered points are the measured values, solid scattered points are from Ref. [16], solid line is the fitting curve with a quadratic polynomial, and the dash line is from the empirical Eq. (6).

It is known that the relative intensity changes of diffraction peaks will take place when a polycrystalline sample presents preferred orientation, namely $I_k=J_k \cdot P_k$. Here, $I_k$ is the intensity of the $k^{th}$ diffraction peak of a preferred orientation sample, and $J_H$ is the intensity of the $k^{th}$ diffraction peak of the corresponding uniform sample without preferred orientation. $P_k$ is defined as the degree of preferred orientation. Assuming the degree of preferred orientation of the strongest diffraction peak is $P_H$, then $P_k = (I_k/I_H) \cdot (J_H/J_k) \cdot P_H$. Where $I_k/I_H$ is the intensity ratio of diffraction peaks from the sample with preferred orientation $H$, and $J_H/J_k$ is also the intensity ratio of the corresponding diffraction peaks but from the uniform sample without preferred orientation, which usually can be obtained from the standard powder diffraction cards. Defining that the summation of all $P_k$ equals to one, namely, $\Sigma_k P_k=1$, then the degree of preferred orientation of a specific orientation $H$ can be obtained.

Finally, the variations of preferred orientation and the total integral intensity of the crystalline phase are obtained from the in-situ XRD measurements for the heating metal indium film as shown in Fig. 8(a) and 8(b), respectively. It can be found that the preferred orientation of the (200) planes is near 99% as the sample temperature was lower than 130 ℃. However, the orientation degree of (200) plane decreases abruptly around 130 ℃, and reaches to zero at 145 ℃. Simultaneously, the orientation degree of other planes, for example (110) and (211) planes, has an obvious increase. The last faded crystalline planes is the (110) planes. Although the (110) reflection was almost persisted until the melting of metal indium, its diffraction intensity was gradually faded away. This result demonstrates that the volume of crystalline regions become small and small with the temperature approaching to the melting point. Fig. 8(b) indicates that the change of the total integral intensity is not monotonically decreasing in the whole heating process. In fact, the total integral intensity come mainly from the (200) reflection, thus the change of total integral intensity reflects mainly the intensity change of (200) reflection. It can be seen that there are two extreme values in the integral intensity. The first extreme value is located around 60 ℃. Probably, the initial heating eliminates partially the internal stress and micro-defect in the metal indium so that the structure became more ordered, leading the first extreme value of the integral intensity. With the temperature increase, the slipping of crystal planes occurs so that some crystal grains are split and the grain boundary is increased, leading the total decrease of the total diffraction intensity. When the sample was further heated to about 120 ℃, the second extreme value of the total integral intensity appears. In this stage, the move of the crystal grains is easier. To decrease the system energy, the adjacent crystal grains tend probably to aggregate together, leading the increase of the integral intensity to form the second extreme value. However, the second extreme value is only persisted for a short temperature region. Afterward, the crystalline structure of metal indium was quickly destroyed by the disintegration of crystal grains and the higher thermal disorder. After 130 ℃, the total integral intensity of the diffraction peaks is faded away until that the metal indium was completely melted at 156.5 ℃.

## 4 Conclusion

The thermal expansion behavior of metal indium has been studied by in-situ XRD measurements with the Mythen detector system. This study demonstrates that the Mythen detector can be well used for in-situ full-profile XRD measurements, including time-resolved XRD experiments. Rietveld refinement has been used to extract the structural parameters of crystalline metal indium. The conclusions can be summarized as follows: (1) From room temperature to melting at 156.6 ℃, there is no detectable phase transform for the tetragonal metal indium. (2) Lattice parameter $a$ and unit-cell volume increase with temperature, while lattice parameter $c$ and the axis ratio of $c/a$ decrease with temperature. The tetragonal unit cell has a tendency to become cubic one with the increase of temperature. (3) The thermal expansion behavior of metal indium is nonlinear. The variations of $a$-axis, $c$-axis, and unit-cell volume with temperature can be described by biquadratic, cubic, and cubic polynomials, respectively.